\def \SAIT #1 #2 {{\em Mem.\ Soc.\ Astron.\ It.\/} {\bf #1}, #2}
\def \MESS #1 #2 {{\em The Messenger\/} {\bf #1}, #2}
\def \ASTRNACH #1 #2 {{\em Astron. Nach.\/} {\bf #1}, #2}
\def \AAP #1 #2 {{\em Astron. Astrophys.\/} {\bf #1}, #2}
\def \AAL #1 #2 {{\em Astron. Astrophys. Lett.\/} {\bf #1}, L#2}
\def \AAR #1 #2 {{\em Astron. Astrophys. Rev.\/} {\bf #1}, #2}
\def \AAS #1 #2 {{\em Astron. Astrophys. Suppl. Ser.\/} {\bf #1}, #2}
\def \AJ #1 #2 {{\em Astron. J.\/} {\bf #1}, #2}
\def \ANNREV #1 #2 {{\em Ann. Rev. Astron. Astrophys.\/} {\bf #1}, #2}
\def \APJ #1 #2 {{\em Astrophys. J.\/} {\bf #1}, #2}
\def \APJL #1 #2 {{\em Astrophys. J. Lett.\/} {\bf #1}, L#2}
\def \APJS #1 #2 {{\em Astrophys. J. Suppl.\/} {\bf #1}, #2}
\def \APSS #1 #2 {{\em Astrophys. Space Sci.\/} {\bf #1}, #2}
\def \ASR #1 #2 {{\em Adv. Space Res.\/} {\bf #1}, #2}
\def \BAIC #1 #2 {{\em Bull. Astron. Inst. Czechosl.\/} {\bf #1}, #2}
\def \JSQRT #1 #2 {{\em J. Quant. Spectrosc. Radiat. Transfer\/} {\bf #1}, #2}
\def \MN #1 #2 {{\em Mon. Not. R. Astr. Soc.\/} {\bf #1}, #2}
\def \MEM #1 #2 {{\em Mem. R. Astr. Soc.\/} {\bf #1}, #2}
\def \PLR #1 #2 {{\em Phys. Lett. Rev.\/} {\bf #1}, #2}
\def \PASJ #1 #2 {{\em Publ. Astron. Soc. Japan\/} {\bf #1}, #2}
\def \PASP #1 #2 {{\em Publ. Astr. Soc. Pacific\/} {\bf #1}, #2}
\def \NAT #1 #2 {{\em Nature\/} {\bf #1}, #2}

\documentstyle{memsait}
\input epsf.sty
\begin{opening}
\title{THE CHEMICAL COMPOSITION OF THE RARE J-TYPE CARBON STARS} 
\author{CARLOS ABIA$^1$, JORDI ISERN$^2$}
\institute{$^1$Dpto. F\'\i sica Te\'orica y del Cosmos, Universidad de Granada, 18071 Granada, Spain\\
$^2$Institut d'Estudis Espacials de Catalunya - CSIC, Barcelona, Spain}
\date{} 
\end{opening}

\begin{document}

\oddpagefooter{}{}{} 
\evenpagefooter{}{}{} 
 
\bigskip
\begin{abstract}

Abundances of  lithium, heavy elements and  carbon isotope ratios have
been   measured  in 12 J-type galactic    carbon  stars. The abundance
analysis shows that  in   these  stars the abundances   of   s-process
elements with respect to the metallicity  are nearly normal. Tc is not
present in most of  them. The Rb abundances, obtained from  the resonance 
7800 {\AA} Rb I line, are surprisingly low, probably due  to strong 
non-LTE effects. Lithium and $^{13}$C are  found to be  enhanced  in all 
the  stars. These  results are used to discuss the origin of J-stars. 

\end{abstract}

\section{Introduction}
Among the C-stars there exists a significant group of stars ($\sim 15\%$) 
named J-type stars  Bouigue (1954) showing strong  $^{13}$C-bearing molecule
absorptions,  which   usually implies   low   $^{12}$C/$^{13}$C ratios
($<15$) (e.g. Ohnaka \& Tsuji 1999). 
The location of J-stars in the AGB phase is far from
clear. In  fact, some authors have  located these stars  in   a
different  evolutive sequence from that   of the ordinary carbon stars
(e.g. Chen \& Kwok 1993), or even outside the AGB phase. Theoretically, it is not 
easy to obtain an AGB star with   the chemical  peculiarities  presented  by  J-stars.  Low
$^{12}$C/$^{13}$C ratios can be obtained in current AGB star models of
M$\geq 4$ M$_\odot$ by the    so-called   hot bottom   burning (HBB). However,    
this mechanism at  the same time destroys $^{12}$C and, in consequence, the  
C/O ratio  in   the  envelope   is  reduced and    the  star again becomes  
O-rich. Thus,  a  fine-tuning of  the parameters  of   the AGB models
(mass, mixing-length, mass-loss rate, metallicity, etc.), that determine
the chemistry of the envelope, is required to obtain a J-star.  
  
The  presence  of strong $\lambda  6708$ {\AA} Li I lines is frequent in  J-stars.  
About 70$\%$ of the galactic J-stars observed are Li-rich (Boffin et al. 1993). Interestingly,  
HBB can simultaneously produce Li-rich  and $^{13}$C-rich AGB  stars in models
with initial  mass M$\geq  4$  M$_\odot$. However, observations indicate  
that the majority of C-stars in  the galaxy are  low-mass objects, M$\leq 2-3$ M$_\odot$ (e.g. 
Claussen et  al. 1987). For this mass  range no  HBB has been 
found in any AGB model. An important consequence of the third dredge-up (TDU) in  
the AGB phase is the enrichment of the envelope with s-process  elements. These elements 
are  believed to be synthesized during the period between thermal pulses via the 
$^{13}$C($\alpha$,n)$^{16}$O reaction as the neutron source. Thus, if J-type stars owe 
their $^{12}$C enhancement to the  operation  of the TDU, {\it they  should also show some  
s-process element enrichment}. 

In this work we  perform  a detailed abundance  analysis of  twelve galactic J-type
carbon  stars using very   high  resolution and signal-to-noise  ratio
echelle spectra. Our  results, together   with   CNO  and Li  abundances
determined in  other studies, are  contrasted with  theoretical stellar
models to find an evolutionary status  for J-type stars.

\section{Observations}

Details on observations, intrumentation used and data reduction can be found
in Abia et al. (1999). Our stars fullfil the criteria for the $^{13}$C abundance by 
Keenan (1993) to be classified as J-stars. However, WZ  Cas shows weaker  
CN and C$_2$ band absorptions than the other J-stars  in the sample.  
It also shows very strong Na  D lines and its  spectrum does not  look  as crowded as the
rest of the J-stars. Indeed, this happens when the  C/O ratio in the atmosphere
is very close to unity, a characteristic which defines a SC-type carbon
star. Since our  C/O estimate in this  star is $\sim 1.01$, we believe
that WZ Cas has to be considered a SC-star rather than a
typical J-star. Furthermore, it   is the most luminous star  in
our sample (M$_{\rm{bol}}\sim -6.44$), which  could indicate that WZ Cas
belongs to a different population (massive) of  C-stars or that it
is in a different evolutionary status (more evolved)  than the rest of
the J-stars   in  our sample.   Less obviously,  the same peculiarities are  
observed  in the spectrum   of WX   Cyg although for this star,  
we do not have a  clear opinion about its spectral type.
To estimate absolute bolometric magnitudes M$_{\rm{bol}}$ of our stars
we   have  used    the empirical   relationships  between
M$_{\rm{bol}}$  and M$_{\rm  K}$-M$_{\rm  V}$ for C-stars obtained  by
Alksnis et al. (1998). To  obtain M$_{\rm K}$ and M$_{\rm  V}$, the K  and 
V average  values (the  stars studied are variable!) quoted in the SIMBAD database 
were used. K and V magnitudes were corrected for interstellar  extinction according 
to  the galactic extinction  model by  Arenou et al. (1992).  Distances  were derived   
from parallax  measurements by HIPPARCOS.  Some parallaxes have
considerable erros, thus  the absolute magnitudes derived have to be 
considered as average values and only indicative.  

\section{Analysis}

For the majority  of the stars  studied here the effective temperature
is derived by   Ohnaka  \& Tsuji  (1999) using the infrared flux method. For 
some stars we   used the $(J-L')_o$ vs. T$_{\rm{eff}}$  calibration
also described by Ohnaka \& Tsuji (1996). The set of models used in  
this analysis was  computed by the Uppsala
group  (see  Eriksson et al.  1984,  for  details).  The input  elemental  abundances
adopted   for  the J-star  models  were  the  solar values,  with the
exception of C, N and  O which were assumed  to be altered relative to
the Sun.   For each star a model  atmosphere was interpolated in  
T$_{\rm{eff}}$ and C/O ratio in this grid. A typical microturbulence velocity for AGB 
stars  $\xi=3$ kms$^{-1}$ was  adopted or taken  from  the literature when  available
Lambert et al. (1986). 
  
Basically,  we have used three atlases for atomic line identification.
These are those by Utsumi (1970) in the  region   between $\lambda  4400-4500$  
and $\lambda 4750-4900$ {\AA}, and by Wallerstein (1989) and Barnbaum et al.
(1994) in  the region, $\lambda\sim 5000-8000$
{\AA}.  
We followed the same criteria as in Abia \& Wallerstein (1998; hereafter Paper I) 
to consider an identification as useful for abundance analysis. We refer the
reader to this paper for details. Unfortunately, very
few lines were found to be useful for analysis depsite several 
hundred  atomic   lines were searched   in each  star.  For some species only  
{\it  one} line was found. Equivalent widths of the lines were measured
with the SPLOT  program  of the  IRAF  package. We  estimate the
error in the equivalent width from the theoretical expression given in Paper I:    
$\Delta\rm{W}(\lambda)=10$ to 35 m{\AA},  according 
to the line intensity and  to the  S/N  of  the  spectrum,   with the  main  
uncertainty  being introduced by the continuum placing. When  possible, gf values were derived
from identification and   equivalent width measurements  in  the Solar
Atlas by Moore et al. (1966), using solar abundances from Anders \& Grevese (1989).
Otherwise, we used the gf-values given in  the VALD database (Piskunov et al. 1995).


\section{Abundance results}

Table 1 shows our abundance results. The Li abundances were  derived by spectrum  
synthesis  and corrected by N-LTE effects  according  to  Abia et al. (1999). From Table 1  
it is clear that all  the stars have unusual  Li abundances  (log $\epsilon$(Li)$\geq 1$). WX 
Cyg and WZ  Cas are certainly super Li-rich stars, although these stars may not 
be J-type stars (see above). Figure 1 shows the correlation of Li
abundances versus $^{12}$C/$^{13}$C ratios found in J- (this work) and
N-type carbon stars (Abia \& Isern 1997). 

This study is the first  detailed search for the presence
of  Tc in  J-type stars. We have used the intercombination Tc line at 
$\lambda  5924.47$  {\AA}.  We followed the same procedure in the analysis 
as in  Paper I.  As there, the $\lambda 5924$ Tc  blend is not well  
reproduced by synthetic spectra.  Thus, we  prefer to be cautious
and record the Tc abundance as  equal-to-or-less-than. In  most of the stars, the best
fit  to the  Tc blend is   compatible  with no-Tc. For these stars we quote a 
{\it no} entry in Table 1, meaning that Tc, very probably, is not present. 
Leaving  apart the upper limits set for WZ Cas and WX Cyg, possible SC-type stars, 
we can conclude that most of J-stars do not show Tc. 

\begin{figure}
\epsfysize=20cm
\epsfxsize=15cm
\epsfbox{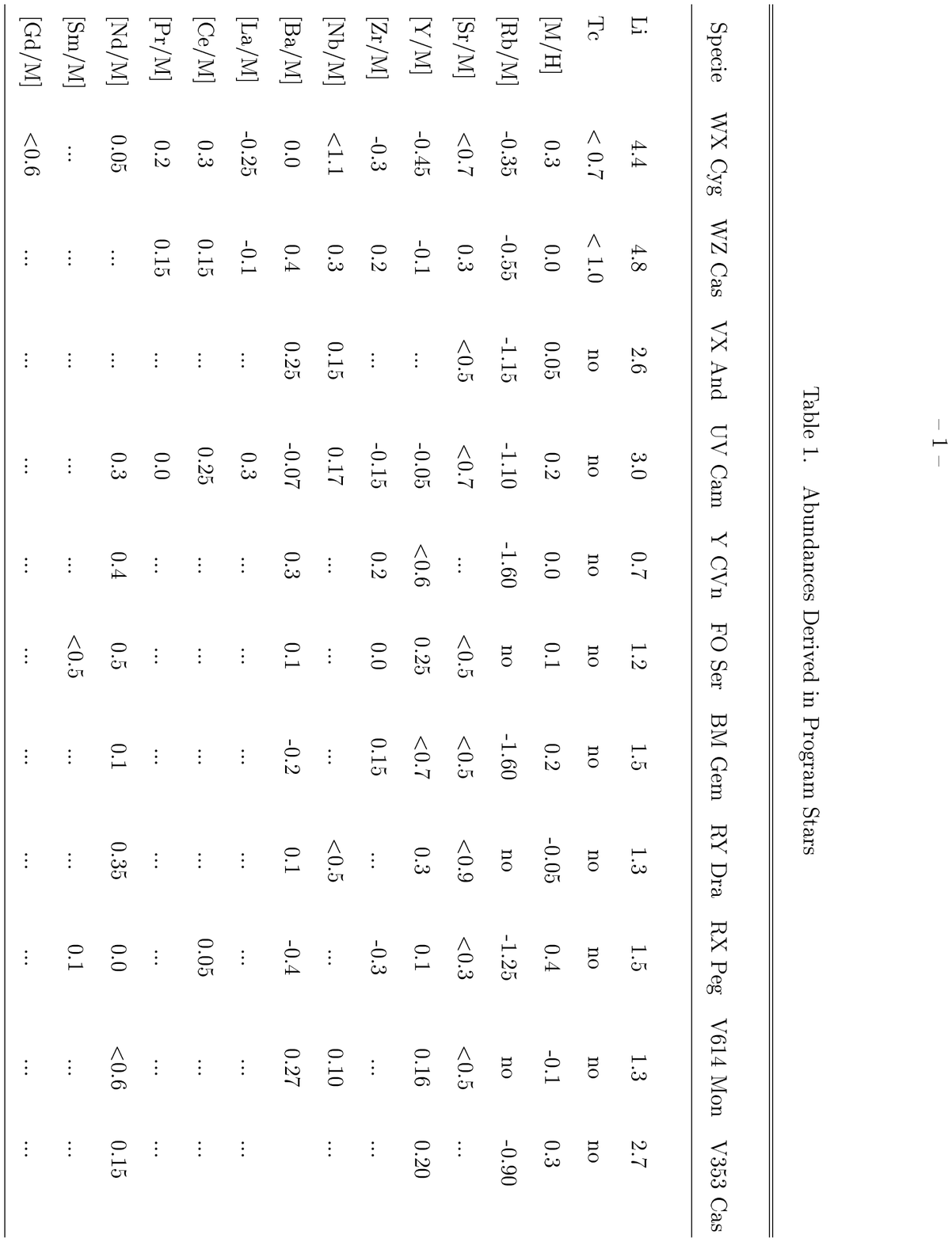}
\caption[h]{}
\end{figure}

\begin{figure}
\epsfysize=6cm 
\epsfxsize=8cm 
\hspace{2.7cm}\epsfbox{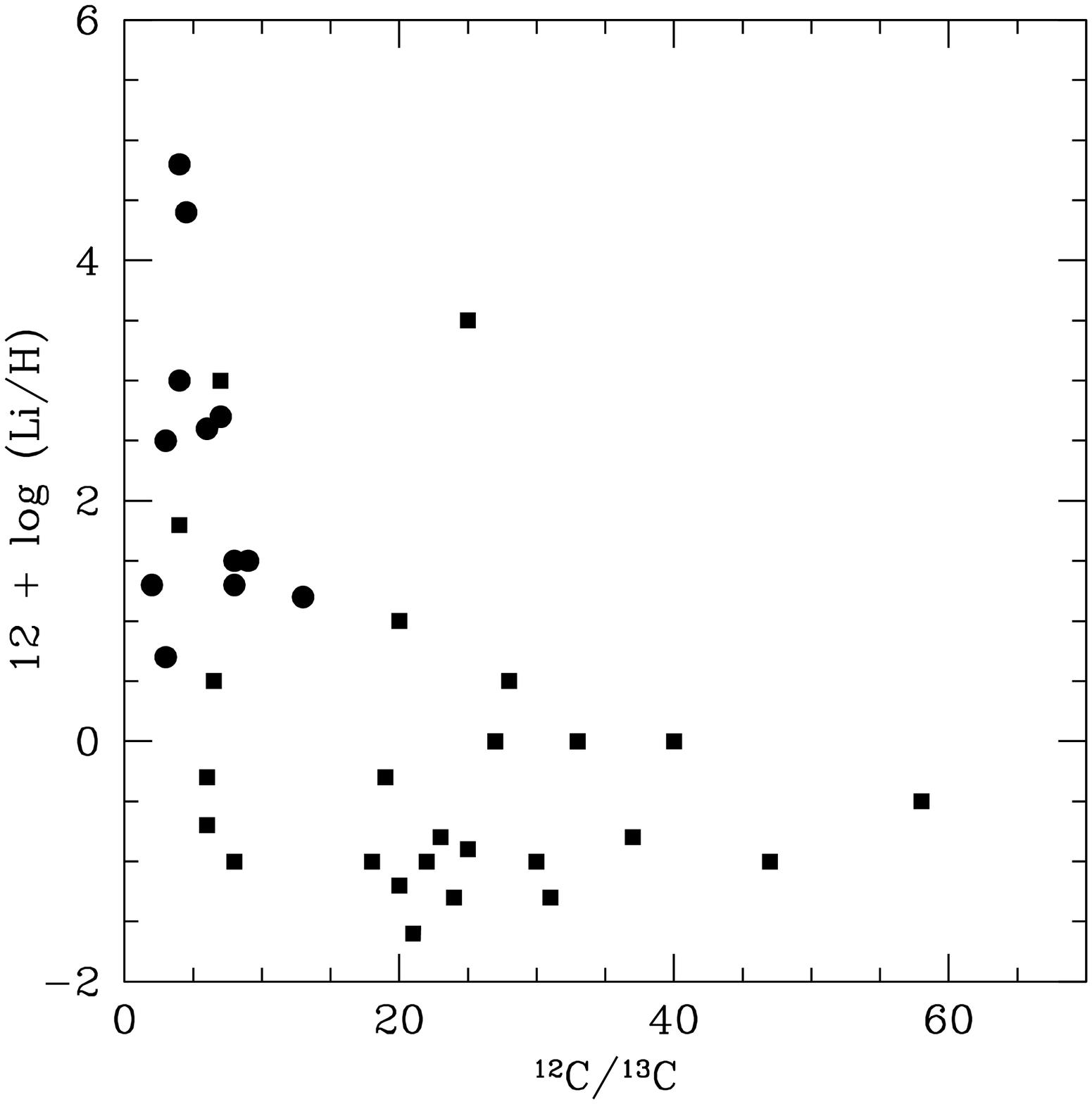} 
\caption[h]{Li abundances vs. $^{12}$C/$^{13}$C ratios in J-stars (circles) in this
work and normal (N) carbon stars (squares) from Abia \& Isern (1997)}
\end{figure}

We have used  the resonance line at $\lambda 7800.23$ {\AA}  to derive Rb 
abundances. We refer again to Paper I for a discussion  of the identification 
of the atomic and molecular  lines contributing to  
the  Rb blend.  Only in three stars (WX Cyg, WZ  Cas and V353 Cas) does the
Rb line  appear clearly as a prominent  absorption in the background
of CN lines.  In the remaining stars  the Rb line  is not distinguished
from the background of lines. Table 1 shows the
Rb  abundance derived  in our  stars   relative  to their  mean
metallicity  [M/H]\footnote{We    adopt  here  the   usual    notation
[X]$\equiv$log(X)$_\star$-log(X)$_\odot$  for  any  abundance quantity
X.}. From  Table 1  it   is apparent  that the   [Rb/M] ratios derived  are
remarkably low. For  some stars the best  fit is  compatible with {\it
no} Rb. Nevertheless, we believe that our  Rb abundances could be, and
in  some   cases  are,   lower  limits probably due to strong N-LTE effects
in the formation of the Rb resonance line in cool C-rich atmospheres
(see Abia \& Isern 1999).

The abundances of metals were derived from the usual method of equivalent width 
measurements and curves of growth calculated in LTE. Ca, V, Fe and Ti abundances 
were used  as a measure of the metallicity of  the stars. The  [M/H]   value  shown  
in  Table  1  is the   mean metallicity obtained from these elements. 
Table 1 also shows the heavy-element abundance ratios respect to the metallicity 
derived in the sample stars. We derive the mean heavy-element enhancement 
[$<h>$/M] in each star. To derive this we did not consider upper limits or
the   uncertain Rb  abundances. The formal error in the heavy-element abundances
shown in Table 1, range from $\pm 0.3-0.6$ dex. Taking this 
error bar into account our results show that   J-type  C-stars  
are  of near   solar  metallicity $\overline{\rm{[M/H]}}=0.12\pm 0.16$, the  
mean heavy   element enhancement  among  the J-stars in the sample being 
[$<h>$/M]$=0.13\pm 0.12$, which is compatible with {\it  non-enrichment}. 

\section{Evolutionary Considerations}

Figure  2  shows the position of  our  J-stars in an observational H-R
diagram,  including some galactic  R-type  and N-type carbon stars with
absolute   magnitudes also derived     from the HIPPARCOS  parallaxes. 
From  this figure, one might  consider
J-stars as  transition objects between   R-stars and N-stars. This  is
reinforced considering the fact that most J-stars are irregular or
semi-irregular variables (very few  Miras  are found among them)  with
not  very  large pulsation periods,  which  is a characteristic of the
less evolved carbon stars.

\begin{figure}
\epsfysize=7cm 
\epsfxsize=8cm 
\hspace{2.7cm}\epsfbox{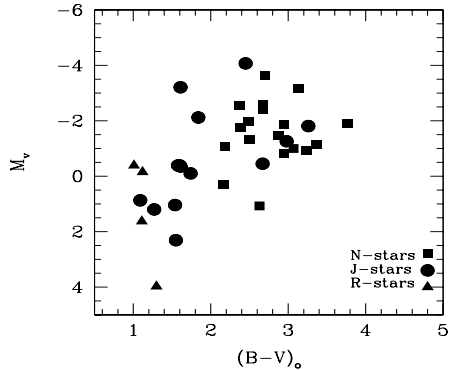} 
\caption[h]{Observational H-R diagram of galactic carbon stars of different
types.}
\end{figure}

Current AGB   models (see Lattanzio, this proceeding) can obtain  C-rich 
envelopes and low carbon  isotope ratios   in stars  with initial mass   
M$\geq 4$ M$_\odot$   through the successive   He-shell  flashes and TDU  episodes
coupled  with  the  operation  of HBB.  These stars can also be,  for a long  
period of time, Li-rich stars. However, the operation of HBB leads to the
transformation of $^{12}$C into $^{14}$N; thus nitrogen is expected to
be enhanced in these stars. The nitrogen  abundances
derived in some J-stars (Lambert et al. 1986) show a normal N/O ratio,
much   lower  than  that expected    on the basis   of  the  CNO cycle
operation in HBB. Theoretical models  can only obtain a C-rich, Li-rich, 
$^{12}$C/$^{13}$C low and N/O$<1$ AGB  star in a very narrow range  
of stellar masses (M$\sim  5$  M$_\odot$), with a specific   metallicity 
(Z$\sim$ Z$_\odot$/3)  and for a very short period  of time ($\leq 10^4$  yr).   
In this context, the number of J-stars expected would be very  low, which is  
in contrast with  the significant number observed. On  the other hand, these 
objects would be fairly luminous  (M$_{\rm{bol}}<-6$), and should present  some
s-process  element enhancement. None  of this is observed.
Wasserburg et al. (1995) have porposed the  existence  of a
non-standard mixing mechanism which might transport material from the bottom
of the  convective envelope into  deeper and hotter regions where  a 
{\it cool  processing} might occur. This hypothetical mixing mechanism has 
been shown to reproduce the CNO isotope  anomalies found in  some  low-mass red  
giants (Boothroyd  \& Sackmann  1999). Under certain conditions,  it  can also create $^7$Li
via the Cameron \& Fowler (1971) mechanism, thus accounting for the
recent discovery  of   surprisingly high  lithium  abundances in  some
low-mass red giants.   Boothroyd    \& Sackmann suggest   that  
this  extra-mixing and  cool    bottom processing could also occur in low-mass  
($\leq 2-3$ M$_\odot$) stars on the early-AGB or just after the onset 
of the helium shell flashes. This point might  be compatible with the suggestion  
(Figure 2) that J-stars are not very  evolved AGB.  In that case, little or no 
s-process element enhancement would be expected, as a  significant number of TDU
episodes are needed. This might also be compatible with the abundance results presented
here. 

Next, we examine  a scenario outside the AGB phase: the mixing at the  He-flash.   
This mechanism has already been proposed to explain   the evolutionary status  of the R-stars
(Dominy 1985). Note that as far  as the chemical composition is concerned,
R-stars and  J-stars   only  differ in  the   presence of   Li  and slightly
lower $^{12}$C/$^{13}$C ratios in the latter. Recently, Deupree \& Wallace (1996) have 
re-examined the He-core flash performing hellium flash calculations of different intensity. 
The authors estimate the surface abundance anomalies produced by He-flashes with different
peak temperatures. They show that the primary material mixed into and
above the H-shell in all cases is $^{12}$C. For peak temperatures
$\sim 9\times 10^8$ K, important $^{12}$C enhancements can be obtained in 
solar metallicity stars, in shuch a way that the star might become a carbon star.
Interestingly, Deupree \& Wallace claim that their flash computations do not produce 
s-process elements while Li production would   require temperatures   not exceeding  
$\sim 5\times 10^7$   K in the processing zone.  This temperature requirement, however,
appear  rather difficult to attain   (Lattanzio, private communication).

Finally, we consider the mass-transfer scenario in a binary
system. It is difficult to explain the  absence  of  s-process  element 
enhancement and  the  C/O ratios in  our stars within this  scenario. In principle, 
the accreted material  must  be extremely   carbon-rich; the donor star should be a
normal C-star with probably enhanced s-nuclei in the envelope. It is easy
to estimate that a C/O$\geq 5$ in the  material accreted is needed by a  {\it typical} 
$\sim 1$ M$_\odot$ red  giant when applying  this scenario to  
explain the C/O$>1$ ratios observed. This  extreme C/O ratio  is  not observed in  
any C-star. Furthermore, it is unlikely that Li could survive during the mass-transfer 
and  posterior mixing. In fact, extrinsic (binary, no Tc) S stars do  
not usually show the Li enhancements found here (Barbuy et al. 1992).  
Nevertheless, a significant number of J-stars ($5\%-10\%$, LLoyd-Evans 1991) 
show a very uniform 9.85 $\mu$m emission which is believed to be due to the
presence of a silicate dust shell. Silicate   emission is   
usually  associated    with O-rich environments, while J-stars are C-rich  objects.
It has  been  suggested   (e.g. LLoyd-Evans 1991) that the
material  expelled from the now carbon   star, starting while it still
had  an  oxygen-rich envelope,  has  accumulated  in a   disc (or common envelope)
around an unseen hypothetical companion. In fact, Barnbaum et al. (1991) found significant radial 
velocity variations in BM Gem and EU And. This result point out to a binary nature for 
these two J-type carbon stars. Although the binary hypothesis
can probably explain the silicate emission in some J-stars, it is difficult to 
explain how binarity can induce  the chemical properties of J-stars. 
Uunfortunately there are not other radial velocity variation studies nor a search 
for ultraviolet excesses (in the hypothesis that the companion is now a white dwarf) 
to test this scenario for all the observed J-stars. 

\section{Conclusions}
Our most important conclusion is that heavy element abundances in J-type
carbon stars are nearly solar with respect to their metallicity. We did not
found Tc in most of the stars. Considering all our
abundance results, it is difficult to find an evolutionary status for
J-stars. Their average luminosity and variability types leads us to consider
these objects as less evolved than normal (N) carbon stars. However,
standard AGB models are unable to explain all their properties. On the contrary, 
the chemical peculiarities of J-stars suggests the existence of a non-standard
mixing mechanism, similar to that proposed in the red giant branch to 
explain anomalous CNO isotopic ratios and Li abundances. This extra-mixing
mechanism, would act preferably in the early AGB phase of low-mass
stars (M$\leq 2-3$ M$_\odot$). Mixing at the He-core flash and the 
binary system hypothesis may well be alternative scenarios, although fine 
tuning is required to explain all the observed characteristics of 
J-stars within these models.  

\acknowledgements

This work was partially supported by grants PB96-1428 and HI1998-0095

\end{document}